# Generation and control of population difference gratings in a three-level hydrogen atomic medium using half-cycle attosecond pulses

**Rostislav Arkhipov**[1,2] , **Mikhail Arkhipov**[1,2] **and Nikolay Rosanov**[1,2]

[1] St. Petersburg State University, 199034 St. Petersburg, Russia

[2] Ioffe Institute, 194021 St. Petersburg, Russia

**E-mail:** arkhipovrostislav@gmail.com, mikhail.v.arkhipov@gmail.com, nnrosanov@mail.ru

Recently, the possibility of the generation and interaction of unipolar half-cycle electromagnetic pulses with quantum systems has been the subject of active research. Such pulses can have many different and interesting applications. They are able to excite quantum systems very fast. Based on the numerical solution of Maxwell-Bloch equations, this paper theoretically studies the possibility of guiding and ultrafast controlling population difference gratings by a sequence of half-cycle attosecond pulses in a three-level resonant medium. The parameters of the model medium (transition frequencies and dipole moments of the transitions) are close to those found in the hydrogen atom. We show the possibility of guiding periodic gratings and dynamic microcavities on different resonance transitions of the medium. In this case, the medium considered is an example of a spatiotemporal photonic crystal. The refractive index varies rapidly in space and time. The possibility of the use of such gratings as Bragg mirrors in ultrafast optics is discussed. The results obtained are of interest in two actively developing areas of modern optics - the physics of half-cycle pulses and space-time photonic crystals.

## Introduction

The study of the generation and interaction of ultrashort electromagnetic pulses with matter has been one of the most important topics in modern optics since the advent of the first lasers [1]. Attosecond pulses are now in active use for ultrafast control of electron dynamics in matter [2-7]. Progress in this field has led to the award of the Nobel Prize in Physics in 2023. This underlines the importance and relevance of this direction in modern physics [6].

The pulses obtained in these studies consist of multiple half-waves, i.e., they are bipolar. If all but one of the half-waves are cut off, we obtain pulses of shortest duration for a fixed spectral range, the so-called unipolar half-cycle pulses [8-13], For pulses of this type, an important characteristic is the electrical area of the pulse. It is defined as the integral of the electric field strength $E$ over time $t$ at a given point in space, $S_E = \int E(t)dt$ [8-13].

In practice, it is difficult to obtain fully unipolar pulses. However, in some cases it is possible to obtain pulses with a shape close to unipolar pulses, which have a pronounced half-wave field [7,14-17]. Such pulses have attracted the attention of researchers because of the peculiarities of their effect on micro-objects - they can be rapidly excited by transmitting a mechanical impulse to a particle in one direction [18-20]. This opens up the possibility of using them in a variety of applications- in ultrafast control of quantum atoms, molecules and nanostructures [18-20], holography with ultrahigh resolution of fast-moving objects etc. See [8-13] for an overview of research in this direction.

The possibility of generating population difference gratings in a resonant medium is one such interesting application. In general, the interference of two or more coherent quasi-monochromatic laser beams is used to create such gratings when the beams are superimposed in the medium [22]. The gratings produced in this way find various applications in optics. For example, they are used as Bragg mirrors in optical fibers [23-27]. However, the possibility of ultrafast control of such gratings - turning them on, turning them off, changing their period, etc. is obviously hampered by the traditional way of creating gratings based on the interference of overlapping monochromatic beams. Extremely short pulses (ESPs) are very well suited for this purpose [28-39].

In a given spectral interval, half-cycle pulses have the shortest duration [8-13]. In the case of half-cycle pulses, direct interference in the conventional sense is impossible. However, in this case it is possible to generate population gratings due to the interference of the electrical areas of the pulses [10,40,41]. On closer examination, the formation of the gratings can be explained by the interference of the polarization waves induced by the previous pulse with those of the following pulse [28-39]. The possibility of studying and controlling gratings with single and half-cycle pulses has been studied previously in various approximations (two-level medium, small field amplitude approximation in a rarefied medium without consideration of propagation effects, etc.) [28-39]. See [33] for a review of the results of early studies.

In [39] it was shown that ESPs can be used to generate and control gratings in a three-level medium with parameters corresponding to atomic rubidium. In [34], the creation of gratings by a sequence of extremely short pulses (ESPs) in a three-level medium of atomic hydrogen was studied. An approximate solution of the Schrödinger equation at small amplitude of the excitation field when perturbation theory is applicable was used in this work. The model of a three-level medium was also used, but the problem was reduced to an optically thin layer, corresponding to a sparse medium. The effects of pulse propagation in an extended medium were not taken into account.

Such approaches are very simplistic. They do not take into account the spatio-temporal dynamics of the polarization in the inversion. This makes the dynamics of the system much more complex. In addition, the formation of gratings by ESPs has also been studied in a two-level medium [28-30].

Sometimes the question of grating formation in a real multilevel medium by ESPs makes researchers skeptical. This casts doubt on some studies of how subcycle pulses interact with matter in the two-level approximation. Furthermore, the coherent propagation of subcycle pulses with matter has mainly been the subject of two-level approximation studies [40-47]. This raises the urgent question of conducting these studies using more realistic multilevel models, considering spatial and temporal field dynamics, polarization and medium inversion.

However, despite the conventional belief about the inappropriateness of the two-level model for such problems, recent studies show qualitatively similar agreement between the results of the two-level approximation and the consideration of a larger number of medium levels [48-50].

We also note that in our case the refractive index of the medium changes rapidly, both in space and time. Our medium is therefore an example of a spatiotemporal photonic crystal (SPC) [51-53]. Such media have been a very interesting subject of research in recent times, see the reviews [54-55]. In addition to their fundamental importance [51-55], they are also of interest for their potential use in frequency transformation of reflected radiation [56], creation of novel laser sources [57], and other uses [55]. Therefore, the issues discussed below may be interesting from the point of view of examples of creating and controlling SPC properties.

**Simple analytical consideration of the grating's formation in multy-level medium**

Let the system be influenced by a pair of half-cycle pulses with an electrical area $S_{E1,2}$ and a delay $\Delta$ of the form

$$E(t) = E_1 \exp[-t^2/\tau_1^2] + E_2 \exp[-(t-\Delta)^2/\tau_2^2]$$

The aim of this paper is to theoretically analyze the possibility of generating and controlling gratings using a sequence of half-cycle attosecond pulses in an extended three-level medium. The theoretical analysis is based on the numerical solution of the Maxwell-Bloch equations for a three-level medium. In this analysis, the pulses did not overlap in the medium. Atomic hydrogen was used as the medium parameter. This work is an extension of previous studies in the two-level and other mentioned above approximations a broadening of its scope.

More complex and unusual polarization and inversion dynamics are shown to be different from the two-level medium. Specifically, the formation of nonharmonic polarization structures, the possibility of forming different types of population gratings at different medium resonance transitions are shown.

It can then be shown that in first-order perturbation theory applied to time dependent Schrodinger equation (when the excitation amplitude is small) the expression for the populations of bound states has the form [10,58-60]:

$$w_{1k} = \frac{d_{1k}^2}{\hbar^2} S_{E1}^2 \exp\left[-\frac{\omega_{1k}^2 \tau_1^2}{2}\right] + \frac{d_{1k}^2}{\hbar^2} S_{E2}^2 \exp\left[-\frac{\omega_{1k}^2 \tau_2^2}{2}\right] + +2\frac{d_{1k}^2}{\hbar^2} S_{E1} S_{E2} \exp[-\omega_{1k}^2(\tau_1^2+\tau_2^2)/4]\cos(\omega_{1k}\Delta)$$

.

Here $\omega_{1k}$ is the transition frequency, $d_{1k}$ is the dipole moment of the transition. This expression is formally similar to the expression for the total intensity of radiation at a given observed point when two monochromatic waves interfere. In this sense, it can be said that the formation of the grating is due to the interference of the pulsed areas [10,58-60].

In the case of an extended but sparse medium, this expression describes the formation of a periodic grating of populations when the medium is exposed to a pair of pulses that travel to meet each other but do not overlap in the medium [28-31,33]. And the delay Δ~z/c (where c is the speed of light in vacuum) is proportional to the time at which the pulse arrives at a given point in the medium with coordinate z. Such an approximation has been used previously to analyze the possibility of grating generation and control [31,34].

The results of calculations of atomic populations with this expression are in good agreement with the results of the numerical solution of the density matrix equations of the two- and three-level medium in the specified sparse medium approximation [31,34,60,61].

However, this approach does not take into account the dynamics of the polarization of the medium in space and time, although it describes a multilevel medium, as mentioned in the introduction. It plays a crucial role in forming the population gratings. It will be taken into account in the next section of this paper.

**Theoretical model and system under consideration**

Therefore, the effects of the propagation of ESPs in an extended medium must be taken into account for a more accurate study. In this case, a joint numerical solution of the Maxwell-Bloch system for the density matrix of a three-level medium and the wave equation for the electric field was performed. This system of equations has the form [62,63]:

$$\frac{\partial}{\partial t}\rho_{21} = -\rho_{21}/T_{21} - -i\omega_{12}\rho_{21} - i\frac{d_{12}}{\hbar}E(\rho_{22}-\rho_{11}) - i\frac{d_{13}}{\hbar}E\rho_{23} + i\frac{d_{23}}{\hbar}E\rho_{31}, \qquad (1)$$

$$\frac{\partial}{\partial t}\rho_{32} = -\rho_{32}/T_{32} - i\omega_{32}\rho_{32} - i\frac{d_{23}}{\hbar}E(\rho_{33}-\rho_{22}) - i\frac{d_{12}}{\hbar}E\rho_{31} + i\frac{d_{13}}{\hbar}E\rho_{21}, \qquad (2)$$

$$\frac{\partial}{\partial t}\rho_{31} = -\rho_{31}/T_{31} - i\omega_{31}\rho_{31} - i\frac{d_{13}}{\hbar}E(\rho_{33}-\rho_{11}) - i\frac{d_{12}}{\hbar}E\rho_{32} + i\frac{d_{23}}{\hbar}E\rho_{21}, \qquad (3)$$

$$\frac{\partial}{\partial t}\rho_{11} = -\frac{\rho_{22}}{T_{22}} + \frac{\rho_{33}}{T_{33}} + i\frac{d_{12}}{\hbar}E(\rho_{21}-\rho_{21}^*) - i\frac{d_{13}}{\hbar}E(\rho_{13}-\rho_{13}^*), \qquad (4)$$

$$\frac{\partial}{\partial t}\rho_{22} = -\rho_{22}/T_{22} - i\frac{d_{12}}{\hbar}E(\rho_{21}-\rho_{21}^*) - i\frac{d_{23}}{\hbar}E(\rho_{23}-\rho_{23}^*), \qquad (5)$$

$$\frac{\partial}{\partial t}\rho_{33} = -\frac{\rho_{33}}{T_{33}} + i\frac{d_{13}}{\hbar}E(\rho_{13}-\rho_{13}^*) + i\frac{d_{23}}{\hbar}E(\rho_{23}-\rho_{23}^*). \qquad (6)$$

$$P(z,t) = 2N_0 d_{12} Re\rho_{12}(z,t) + 2N_0 d_{13} Re\rho_{13}(z,t) + 2N_0 d_{12} Re\rho_{12}(z,t) + 2N_0 d_{23} Re\rho_{32}(z,t). \qquad (7)$$

$$\frac{\partial^2 E(z,t)}{\partial z^2} - \frac{1}{c^2}\frac{\partial^2 E(z,t)}{\partial t^2} = \frac{4\pi}{c^2}\frac{\partial^2 P(z,t)}{\partial t^2}. \qquad (8)$$

Here $\rho_{11}$, $\rho_{22}$, $\rho_{33}$ are the populations of the 1st, 2nd and 3rd states of the atom respectively, $\rho_{21}$, $\rho_{32}$, $\rho_{31}$ are the non-diagonal elements of the density matrix determining the dynamics of the polarization of the medium, $\omega_{12}$, $\omega_{32}$, $\omega_{31}$ are the transition frequencies of the medium and $d_{12}$, $d_{13}$, $d_{23}$ are the dipole moments of these transitions, $\hbar$ is the reduced Planck constant.

The equations also include $T_{ik}$ relaxation terms.

The medium parameters (transition frequencies and dipole moments of the transitions) have been chosen to be close to those realized for the hydrogen atom [64]. They are given in the table.

**Table. Parameters of the model and excitation pulses used in calculations**

| Parameter | Value |
|---|---|
| Transition frequency 12 (corresponding wavelength) | $\omega_{12} = 1.55 \cdot 10^{16}$ rad/s ($\lambda_{12} = \lambda_0 = 121.6$ nm) |
| Dipole moment of transition 12 | $d_{12} = 3.27$ D |
| Transition frequency 13 в атоме водорода (corresponding wavelength | $\omega_{13} = 1.84 \cdot 10^{16}$ rad/s ($\lambda_{13} = 102.6$ nm) |
| Dipole moment of the transition 13 | $d_{13} = 1.31$ D |
| Transition frequency 23 (wavelength) | $\omega_{23} = 2.87 \cdot 10^{15}$ rad/s ($\lambda_{23} = 656.6$ nm) |
| Dipole moment of transition 23 | $d_{23} = 12.6$ D |
| Atomic concentration | $N_0 = 10^{14} cm^{-3}$ |
| Pulse amplitude | $E_{01} = E_{02} = 45 \cdot 10^7$ V/cm |
| Pulse duration | $\tau = 100$ as |
| Relaxation times $T_{ik}$ | $T_{ik} = 1$ ns |

As initial conditions at the initial time, a pair of half-cycle Gaussian pulses was sent from the extremes of the integration region into the medium

$$E(z = 0, t) = E_{01} e^{-\frac{(t-\Delta_1)^2}{\tau^2}}, \quad (9)$$

$$E(z = L, t) = E_{02} e^{-\frac{(t-\Delta_2)^2}{\tau^2}}. \quad (10)$$

Delays $\Delta_1 = 2.5\tau$ и $\Delta_2 = 29\tau$ were chosen in such a way that the pulses did not overlap simultaneously in the medium. The system of equations (1)-(8) has been solved numerically. The density matrix equations (1)-(6) were solved by the 4th order Runge-Kutta method. The solution of the wave equation (8) was by the finite difference method.

The total length of the computational domain is $L = 12\lambda_0$. The medium was situated between the points $z_1 = 4\lambda_0$ and $z_2 = 8\lambda_0$. In the calculations, the pulses reached the boundary of the integration domain. They were reflected from it and returned to the medium, but did not overlap. In Fig. 1, the movement of the pulses is shown schematically by arrows. The numbers denote their number.

# Results of the numerical simulations and discussions

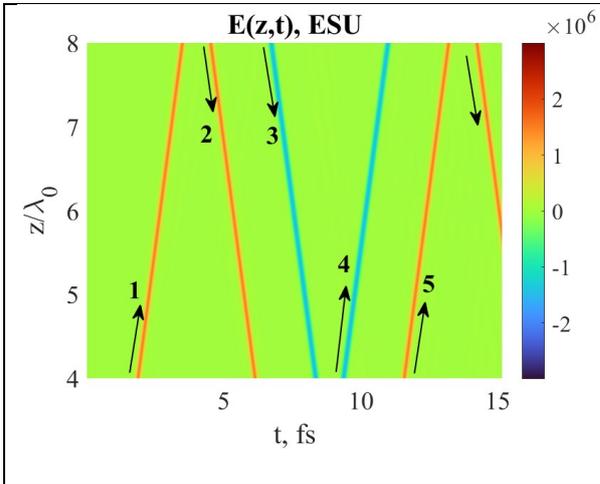

**Fig. 1.** Scheme of the propagation of electric field pulses in the medium

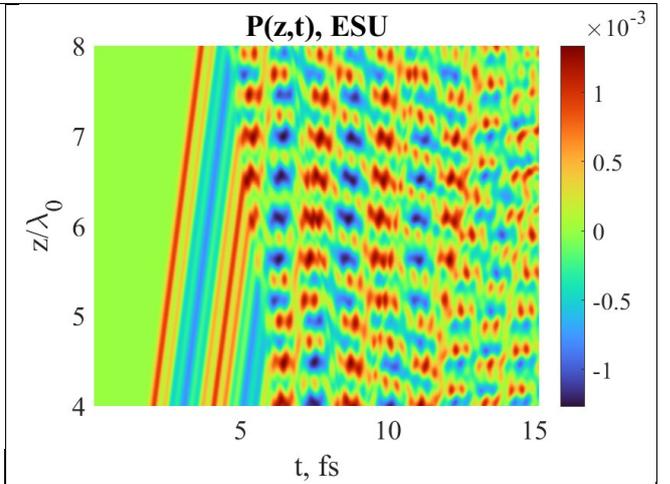

Fig. 2. Spatial and temporal dynamics of medium polarization $P(z,t)$.

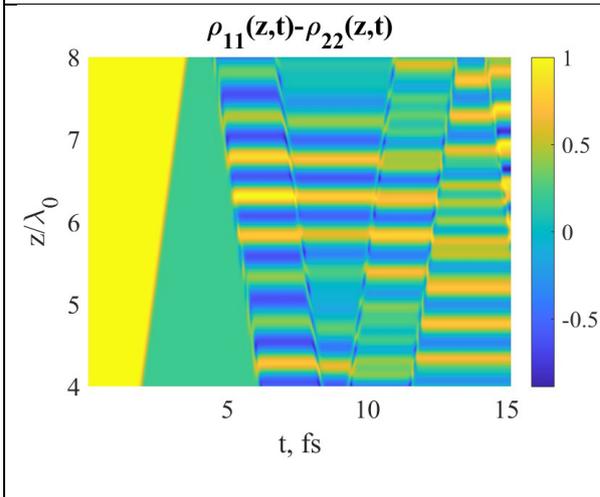

Fig.3. Spatial and temporal dynamics of medium population difference $\rho_{11} - \rho_{22}$

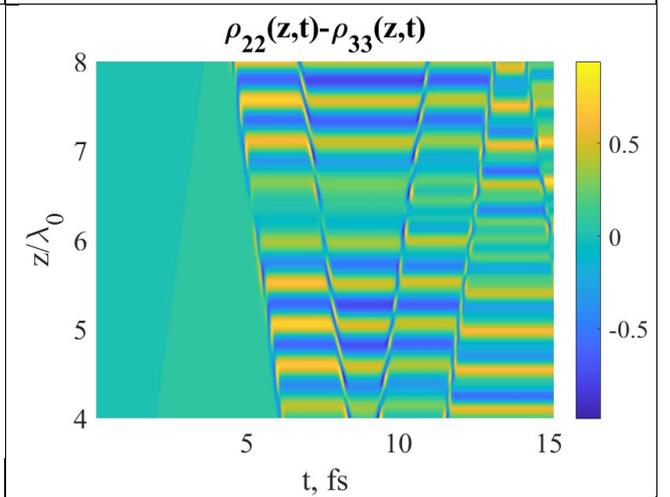

Fig.4. Spatial and temporal dynamics of medium population difference $\rho_{22} - \rho_{33}$

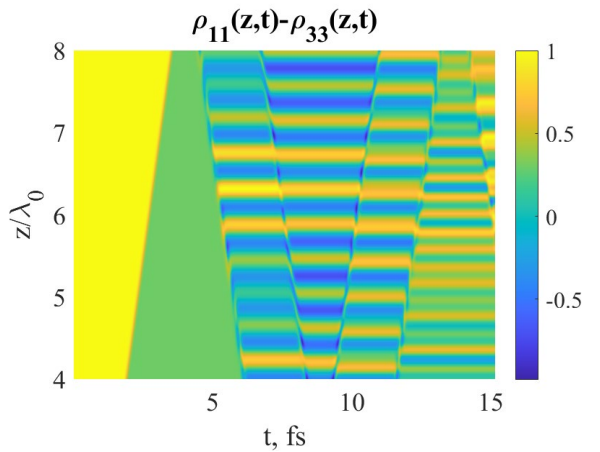

Fig.5. Spatial and temporal dynamics of medium population difference $\rho_{11} - \rho_{33}$

The results of the numerical integration of the system of equations (1)-(8) with the parameters given in the table are shown in Fig.2-5. Fig.2 illustrates the spatial and temporal dynamics of the medium polarization $P(z,t)$, Fig.3-4 the population differences at the transitions 1-2 and 2-3.

From Fig.3 we can see that at transition 1-2 there is a grating formation after pulse 2. However, this grating is not strictly harmonic. The polarization of the medium oscillates not only at the frequency of transition 1-2, but also at the frequencies of other transitions due to the presence of the third level. Figure 2 shows these oscillations. The amplitude of the grating in space is thus modulated.

We can also see from Fig. 3 that the third and subsequent pulses change their parameters - shifting in space, erasing, reducing the depth of modulation, etc. This distinguishes our case from that considered in Rubidium, ref. [39]. In those calculations, the shape of the resulting lattices was closer to the harmonic, as in the two-level medium.

However, at transition 1-3, see Fig. 5, structures appear that are closer in form to harmonic structures. A more interesting situation occurs at the transition 2-3, see Fig. 3. In this case, a microcavity with lateral Bragg mirrors is formed near the point $z = 6.5\lambda_0$. Near this point, the population difference has a constant value. A periodic grating of populations - the Bragg grating - appears on the sides. Similar microcavities appeared when unipolar rectangular pulses collided in a two-level medium [35-38].

Thus, as sometimes assumed, the effect of population gratings predicted in a two-level medium does not disappear when additional levels of the medium are taken into account. The inclusion of the additional levels leads to a change in the form of the gratings, which may be different from the harmonic form. However, the harmonic form is possible as well [39]. Thus, the consideration of other levels of the medium results in a more various dynamics of the system.

**Estimation of Bragg grating reflection coefficient**

In optical fibers [23-27], such gratings can be used as Bragg mirrors. To estimate the reflection coefficient of the grating, we assume that the grating is created in a waveguide, i.e. there is a modulation of the refractive index in the fiber according to the expression (for simplicity, we assume a harmonic dependence):

$$n = n_0 + \Delta n \cos\frac{2\pi}{\Lambda}z,$$

$n_0$ – average refractive index $\Delta n$ – modulation amplitude, $\Lambda$ – period of the grating. The reflection coefficient $R(l,\lambda)$ of the grating can be estimated by means of the expression [23,24,26]

$$R(l,\lambda) = \frac{\Omega^2(\sinh sL)^2}{\Delta k^2(\sinh sL)^2 + s^2(\cosh sL)^2}.$$

here, $l$ is the grating length, $\lambda$ is the wavelength, $s = \sqrt{\Omega^2 - \Delta k^2}$, $\Omega \cong \pi\Delta n/\lambda$, $\Delta k = k - \pi/\Lambda$ –detuning, $k = 2\pi n_0/\lambda$ is the propagation constant.

The typical dependence of grating reflection coefficient R on radiation wavelength $\lambda$ for different values of $\Delta n$ is shown in Figure 6. In this example, $\Lambda = \lambda_0/2$, $l = 5\lambda_0$, $n_0 = 1$.

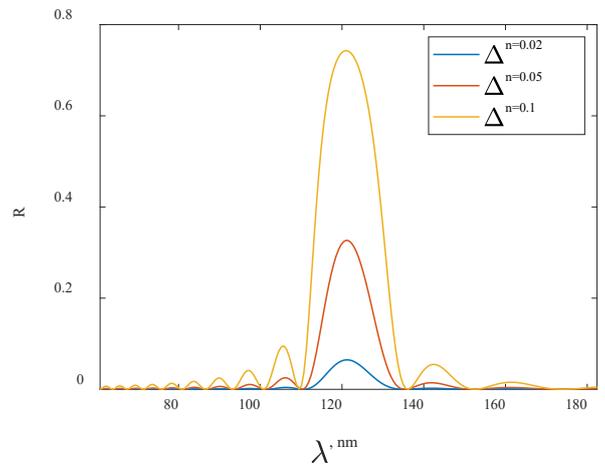

**Fig. 6.** Dependence of Bragg grating reflection coefficient $R(l,\lambda)$ on wavelength $\lambda$

for different refractive index modulation depths $\Delta n$. $\Lambda = \lambda_0/2$, $l = 5\lambda_0$.

Obviously, increasing the modulation depth, which can be varied with the amplitude and duration of the incident pulses, leads to an increase in reflectivity. Therefore, the induced gratings can be used as Bragg mirrors with rapidly varying parameters and as highly reflective mirrors in dynamic microcavities

**Conclusions**

In this work we have demonstrated the possibility of generating and ultrafast controlling population difference gratings in a three-level medium by a sequence of half-cycle attosecond pulses, based on the numerical solution of the system of Maxwell-Bloch equations. The parameters of the medium are found to be close in value to those of the atomic hydrogen. The pulses do not overlap with each other in the medium. It is shown that considering additional levels of the medium also leads to the formation of gratings. However, the dynamics of the system may differ from the case of a two-level medium. This extends the applicability of previous results using two-level and other approximations.

It is shown that the shape of the gratings can, in general, differ from the purely harmonic one previously observed in the two-level medium. Also, the shapes of these gratings generated at different resonance transitions of the medium are very different from each other. It is almost harmonic at some transitions. At others, it is possible to form dynamic microcavities. This has previously been demonstrated in a two-level medium. The reflecting coefficient of Bragg mirrors based on these gratings has been estimated.

These studies open new avenues of research in the optics of unipolar half-cycle pulses [8-13] and in the study of media with rapidly changing parameters - space-time photonic crystals [51-57].

**Funding of the work**

The research was financially supported by Russian Science foundation, project 23-12-00012.

.